\documentclass[12pt,notoc]{JHEP3}

\usepackage{amsmath,amssymb,euscript,array,cite,mathrsfs}

\def\newf{{\cal F}}

\newcommand{\startappendix}{
\setcounter{section}{0}
\renewcommand{\thesection}{\Alph{section}}}
\newcommand{\Appendix}[1]{
\refstepcounter{section}
\begin{flushleft}
{\large\bf Appendix \thesection: #1}
\end{flushleft}}
\usepackage{epsfig}

\def\C{\Gamma}

\newcommand{\sech}{\operatorname{sech}}

\newcommand{\Tr}{\operatorname{Tr}}

\def\B0{{\boldsymbol 0}}
\def\BF{{\boldsymbol F}}
\def\BK{{\boldsymbol K}}
\def\BX{{\boldsymbol X}}
\def\BZ{{\boldsymbol Z}}

\def\BOmega{{\boldsymbol\Omega}}
\def\Bvarpi{{\boldsymbol\varpi}}
\def\Bpi{{\boldsymbol\pi}}
\def\CP{{\mathbb C}P}
\def\RP{{\mathbb R}P}

\def\Tr{{\rm Tr}}

\def\C{{\mathbb C}}

\def\ee{\boldsymbol{e}}

\newcommand{\BH}{\boldsymbol{H}}

\def\Dbarslash{\,\,{\raise.15ex\hbox{/}\mkern-12mu {\bar D}}}
\def\Dslash{\,\,{\raise.15ex\hbox{/}\mkern-12mu D}}
\def\delslash{\,\,{\raise.15ex\hbox{/}\mkern-9mu \partial}}
\def\delbarslash{\,\,{\raise.15ex\hbox{/}\mkern-9mu {\bar\partial}}}

\def\LAG{\mathscr{L}}

\newcommand{\MAT}[1]{\begin{pmatrix} #1\end{pmatrix}}
\newcommand{\EQ}[1]{\begin{equation} #1 \end{equation}}

\newcommand{\SP}[1]{\begin{equation}\begin{split} #1
\end{split}\end{equation}}

\title{A New and Elementary $\boldsymbol{\CP^n}$ Dyonic Magnon}
\author{Timothy J. Hollowood\\
Department of Physics,\\ University of Wales Swansea,\\
Swansea, SA2 8PP, UK.\\
E-mail: \email{t.hollowood@swansea.ac.uk}}

\author{and J.~Luis Miramontes\\
Departamento de F\'\i sica de Part\'\i culas and IGFAE,\\
Universidad
de Santiago de Compostela\\ 15782 Santiago de Compostela, Spain\\
E-mail: \email{jluis.miramontes@usc.es}}

\abstract{
We show that the dressing transformation method produces a new type
of dyonic $\CP^n$ magnon in terms of which all the other known 
solutions are either composites or arise as special limits. In particular, this
includes the embedding of Dorey's dyonic magnon via
$\RP^3\subset \CP^n$. We also show how to generate 
Dorey's dyonic magnon directly in the $S^n$ sigma model via the
dressing method without resorting to the isomorphism with the $SU(2)$
principle chiral model when $n=3$.
The new dyon is shown to be either a charged dyon or topological kink
of the related symmetric-space sine-Gordon theories associated to
$\CP^n$ and in this sense is a direct generalization of the soliton of the
complex sine-Gordon theory.
}

\begin{document}

\section{Introduction}

Motivated by the investigation of the AdS/CFT correspondence for
$AdS_4\times \CP^3$~\cite{NewDuality}, the $\CP^n$ giant magnons have
been recently discussed in some
detail~\cite{Gaiotto:2008cg,Grignani:2008is,Grignani:2008te,Astolfi:2008ji,Abbott:2008qd,Kalousios:2009mp,Suzuki:2009sc,Abbott:2009um}. 
Similarly to their $S^n$ counterparts, they are soliton solutions to
the equations of motion of a Pohlmeyer reduced sigma model with target
space $\CP^n$. 

So far, four different kinds of $\CP^n$ giant magnons have been
described in the literature. Two of them are obtained by embedding the
original $S^2$ Hofman-Maldacena giant magnon~\cite{Hofman:2006xt}  in
two distinct subspaces; namely, $\CP^1 \subset
\CP^n$~\cite{Gaiotto:2008cg}, and ${\mathbb R}P^2 \subset\CP^n$
($n\geq2$)~\cite{Grignani:2008is}. They have one parameter, and carry
a single non-vanishing conserved charge (angular momentum). The third
one is obtained by embedding Dorey's dyonic $S^3$ giant
magnon~\cite{DyonicGM} via ${\mathbb R}P^3 \subset\CP^n$
($n\geq3$)~\cite{Abbott:2008qd}, and it is a two-parameter (dyonic)
generalization of the ${\mathbb R}P^2$ magnon that carries two
conserved charges. 
The fourth $\CP^n$ magnon was recently constructed by the present 
authors in~\cite{Hollowood:2009tw} using the dressing
method (see also~\cite{Kalousios:2009mp,Suzuki:2009sc}). It has two parameters and takes values in a $\CP^2$
subspace but carries only a single conserved charge.  
The relationship between all these (sigma model) giant magnons and
those obtained from algebraic curves is discussed
in~\cite{Abbott:2009um}. 

The purpose of this note is to argue that all those magnons can be
built out of a new type of $\CP^n$ dyonic magnon that we construct
using the dressing transformation method. 
The existence of additional $\CP^n$ dyonic solutions was conjectured
in~\cite{Hollowood:2009tw} as a consequence of the general form of the
metric on the moduli space of internal collective coordinates.  
Moreover, the main features of a solution of precisely the type of the
new dyon were discussed in~\cite{Abbott:2009um}, although its explicit
form was only found there for a particular value of the parameters.  
Our conclusion will be that the dressing method can produce all known
solutions, either as composites of the new dyonic magnon, or as special
limits of a single one. In particular, we will argue that the embedding
of Dorey's dyonic magnon in a subspace $\RP^3\subset\CP^n$ is a
composite configuration of two of the
new dyonic magnons and has internal moduli corresponding to separating
the constituents. As a by-product we are able to show how Dorey's magnon
can be constructed in the $S^n$ theory directly by using the dressing
method. In this context, it is a fundamental object which cannot be
``pulled apart''. 

One of the main results of \cite{Hollowood:2009tw} was that the
dressing method naturally produces magnon solutions in the original
sigma model and at the same time the associated solution---the
``solitonic avatar''---in the associated Symmetric-Space Sine-Gordon
(SSSG) theory. This fact allows us to also investigate the nature of
the avatar of the new dyonic magnon.

The paper is organized as follows. In Section~\ref{SMsection} we
review the construction of the $\CP^n$ sigma model and in
Section~\ref{DT} we describe how to impose the Pohlmeyer reduction and
then how to construct solutions using the dressing transformation. In
Section~\ref{mag} we construct the simplest kinds of solution and
in particular previously over-looked solutions that are the new
elementary dyonic magnons. In Section~\ref{avatar} we consider the new
solutions from the point-of-view of the associated SSSG system of
equations. Finally in Section~\ref{Mother} we show that all the known
solutions can be recovered from the new dyons.

\section{The $\CP^n$ Sigma Model}
\label{SMsection}

The $2n$-dimensional complex projective space 
\EQ{
\CP^n = \frac{\C^{n+1}}{\BZ\sim \lambda \BZ}  \simeq \frac{SU(n+1)}{U(n)}\>,
\label{SS}
}
where $\BZ$ is a complex $n+1$~dimensional vector and $\lambda\in\C$, is a compact symmetric space $F/G$ specified by the involution
\EQ{
\sigma_-(f)= \theta f \theta\>,
}
where
\EQ{
f\in SU(n+1)\>, \qquad \theta= \text{diag}(-1,1,\ldots,1)\>.
}
Acting on ${\mathfrak f}$, which is the Lie algebra of $F=SU(n+1)$, it gives rise to the canonical decomposition
\EQ{
{\mathfrak f} = {\mathfrak g} \oplus {\mathfrak p}
\quad \text{with} \quad 
[{\mathfrak g},{\mathfrak g}]\subset {\mathfrak g}\>, 
\quad [{\mathfrak g},{\mathfrak p}]\subset 
{\mathfrak p}\>, 
\quad [{\mathfrak p},{\mathfrak p}]\subset {\mathfrak g}\>,
\label{CanonicalDec}
}
where, using the fundamental representation of $SU(n+1)$, the form of the elements $r\in{\mathfrak g}$ and $k\in{\mathfrak p}$ is
\EQ{
r=\left(\begin{array}{cccc}in\phi &0&\cdots&0\\
0&&&\\\vdots&& -i\phi I_{n\times n} + {\cal M}\\
0&&&
\end{array} \right)\>, \qquad
k=\left(\begin{array}{cccc}0&v_1&\cdots&v_{n}\\
-v_1^\ast&0&\cdots&0\\\vdots&\vdots&\cdots&\vdots\\
-v_n^\ast&0&\cdots&0
\end{array} \right)\>,
}
with ${\cal M}$ a $n\times n$ anti-Hermitian matrix. Notice that ${\mathfrak g}$ is the Lie algebra of $G=U(n)$.

Following the approach of~\cite{Hollowood:2009tw,Eichenherr:1979hz}, the sigma model with target space $\CP^n$ can be formulated in terms of a $SU(n+1)$-valued field $\newf$ subject to the constraint
\EQ{
\sigma_-(\newf)=\theta \newf \theta = \newf^{-1}\>.
\label{Cond}
}
The map from the space $\CP^n$ into this field is given by
\EQ{
\newf = \theta \Big( I - 2\> \frac{\BZ \BZ^\dagger}{|\BZ|^2} \Bigr)\>,
}
where $\BZ$ is a complex $n+1$ dimensional vector whose components are the complex projective (embedding) coordinates.
Then, the Lagrangian of the sigma model is 
\EQ{
\LAG= -\Tr \big( {\cal J}_\mu {\cal J}^\mu\big)\quad \text{with}
\quad {\cal J}_\mu=\partial_\mu \newf \newf^{-1}\>,
\label{Lag}
}
whose equations-of-motion of~\eqref{Lag} are
\EQ{
\partial_\mu {\cal J}^\mu=0\>.
\label{p1}
}
They exhibit that $ {\cal J}_\mu$ is the conserved current corresponding to the
global symmetry transformation\footnote{The Lagrangian density~\eqref{Lag} is invariant under the global transformations $\newf\to U \newf V$ for any $U,V \in SU(n+1)$. However, this symmetry is reduced by the constraint~\eqref{Cond} so that the $\CP^n$ sigma model is invariant only under~\eqref{Global}.
}
\EQ{
\newf\to U \newf \sigma_-(U^{-1})\>,\qquad
U\in SU(n+1)\>,
\label{Global}
}
which gives rise to the conserved Noether charge
\EQ{
{\cal Q}_L = \int_{-\infty}^{+\infty} \partial_0\newf \newf^{-1}\>.
}

\section{Pohlmeyer Reduction and Dressing Transformations}
\label{DT}

``Giant magnon'' is the name given to a soliton of the Pohlmeyer reduced sigma model in the context of string theory. For the $\CP^n$ sigma model, the Pohlmeyer reduction involves imposing the conditions~\cite{Hollowood:2009tw,Miramontes:2008wt}\footnote{In our notation, $x_\pm = t\pm x$ and $\partial_\pm =\frac{1}{2}(\partial_t\pm\partial_x)$.}
\EQ{
\partial_\pm \newf \newf^{-1} = f_\pm \Lambda f_\pm^{-1}\>,
\label{PohlConst}
}
where $f_\pm\in SU(n+1)$ and
\EQ{
\Lambda=\left(\begin{array}{cc|c}0&-1&\B0\\1&0&\B0\\\hline \B0&\B0&\B0\end{array}\right)\>.
}

Pohlmeyer reduction gives rise to an associated relativistic integrable system that is a
generalization of the sine-Gordon
theory~\cite{Pohlmeyer:1975nb}. These are the SSSG theories, and giant
magnons can be mapped into the soliton solutions of their
equations-of-motion. These latter equations are 
\EQ{
\big[\partial_++\gamma^{-1}\partial_+\gamma+\gamma^{-1}A_+^{(L)}\gamma-\frac{1}{2}\Lambda\>,\;
\partial_-+A_-^{(R)}-\frac{1}{2}\gamma^{-1}\Lambda\gamma\big]=0\>,
\label{www1}
}
where
\EQ{
\gamma=f_-^{-1}f_+
\label{xxd1}
}
is the SSSG group field, which takes values in $G\subset F$. The
quantities $A^{(L)}_+$ and 
$A^{(R)}_-$ can be interpreted as components of gauge fields taking
values in ${\mathfrak h}$, the Lie algebra of $H\subset G$, which is
the subgroup of elements that commute with $\Lambda$. They are given
by 
\SP{
A_+^{(L)}&=f_-^{-1}\partial_+f_-
-\frac{1}{2}\gamma \Lambda\gamma^{-1}\ ,\\
A_-^{(R)}&=f_+^{-1}\partial_-f_+ -\frac{1}{2} \gamma^{-1}\Lambda\gamma\ .
\label{laa}
}
In the present case, $F=SU(n+1)$, $G=U(n)$ and $H=U(n-1)$.

We will be interested in the $\CP^n$ magnons constructed using the
dressing transformation method~\cite{Dressing,Harnad:1983we}, which
was shown to be consistent with the Pohlmeyer reduction
in~\cite{Hollowood:2009tw}. The procedure begins by identifying a
``vacuum'' solution which, in the present context, will be the
simplest one which naturally satisfies the
constraints~\eqref{PohlConst} with $f_\pm=I$. It corresponds to 
\EQ{
\newf_0=\left(\begin{array}{cc|c}\cos 2t&-\sin 2t&\B0\\+\sin2t&\cos2t&\B0\\\hline \B0&\B0&\boldsymbol{I}\end{array}\right)\>, \qquad
\BZ_0= {\bf e}_1 \cos t -  {\bf e}_2 \sin t\>,
\label{VaccumField}
}
where $\{{\bf e}_1, \ldots,{\bf e}_{n+1}\}$ is a set of real orthonormal vectors in $\C^{n+1}$ and we have highlighted the $2\times2$ subspace associated to ${\bf e}_1$ and ${\bf e}_2$.

The dressing transformation method makes use of the associated linear system
\EQ{
\partial_\pm \Psi(x;\lambda) =\frac{\partial_\pm \newf \newf^{-1}}{1\pm\lambda}\Psi(x;\lambda)\>, \qquad
\Psi(x;\infty)=I
\>, \qquad \newf(x)= \Psi(x;0)\>,
\label{LinearP}
}
whose integrability conditions are equivalent to the equations of motion of the sigma model. For $\CP^n$, the solutions $\Psi(x;\lambda)$ have to satisfy the two conditions
\EQ{
\Psi^{-1}(x;\lambda) = \Psi^\dagger(x;\lambda^\ast) \>,\qquad
\Psi(x;1/\lambda)=\newf\theta\Psi(x;\lambda)\theta \>,
\label{CondPsi}
}
which ensure that $\newf^{-1}=\newf^\dagger$ and that the constraint~\eqref{Cond} is satisfied. Then, the dressing transformation involves constructing a new solution $\Psi$ of the linear system of the form
\EQ{
\Psi(x;\lambda)= \chi(x;\lambda) \Psi_0(x;\lambda)
\label{DressingT}
}
in terms of an old one, which in our case correspond to the vacuum solution in~\eqref{VaccumField}:
\EQ{
\Psi_0(x;\lambda)=\exp\Big[ \Big(\frac{x_+}{1+\lambda} +\frac{x_-}{1-\lambda}\Big) \Lambda\Big]\>.
}

Following~\cite{Harnad:1983we}, the general form of the ``dressing factor'' is
\EQ{
\chi(\lambda)= 1+ \sum_i \frac{Q_i}{\lambda-\lambda_i}\>, \qquad
\chi^{-1}(\lambda)= 1+ \sum_i \frac{R_i}{\lambda-\mu_i}\>,
}
where the residues are matrices of the form
\EQ{
Q_i= \BX_i \BF_i^\dagger\>, \qquad
R_i= \BH_i \BK_i^\dagger
}
for vectors $\BX_i$, $\BF_i$, $\BH_i$, and $\BK_i$. For $\CP^n$, they are given by
\SP{
&
\BX_i \Gamma_{ij}=\BH_j\>, \qquad
\BK_i \big(\Gamma^\dagger\big)_{ij} = -\BF_j\>, \qquad
\Gamma_{ij}=\frac{\BF_i^\dagger \BH_j}{\lambda_i-\mu_j}\>,\\[5pt]
&
\BF_i = \Psi_0(\lambda_i^\ast) \Bvarpi_i\>, \qquad
\BH_i = \Psi_0(\mu_i) \Bpi_i\>,
}
where $\Bvarpi_i$ and $\Bpi_i$ are complex constant $n+1$~dimensional vectors. The allowed number of poles and their positions are constrained by the conditions~\eqref{CondPsi}. They imply that $\mu_i=\lambda_i^\ast$ and, moreover, that the poles $\{\lambda_i\}$ must come in pairs $(\lambda_i,\lambda_{i+1}=1/\lambda_i)$. In addition, $\Bpi_i=\Bvarpi_i$ and, for each pair,
\EQ{
\Bvarpi_{i+1}=\theta\Bvarpi_i\>.
}
In~\cite{Hollowood:2009tw}, it was shown that the value of the $SU(n+1)$ charge carried by these ``dressed'' solutions, relative to the vacuum solution, can be easily calculated in terms of the asymptotic values of the residues by means of
\EQ{
\Delta{\cal Q}_L = \sum_i Q_i \Big|_{x=+\infty} -\sum_i Q_i \Big|_{x=-\infty}\>.
\label{Charges}
}

One the main results of \cite{Hollowood:2009tw} is that the dressing
transformation not only produces the magnon solutions but the
associated solitons, the ``avatars'', of the related Symmetric Space
Sine-Gordon (SSSG) equation.
They are given by
\EQ{
\gamma={\cal F}_0^{-1/2}\chi(+1)^{-1}\chi(-1){\cal
  F}_0^{1/2}\ ,
\label{xxd}
}
with $A^{(L)}_+=A^{(R)}_-=0$. The group field also satisfies the
constraints
\EQ{
\gamma^{-1}\partial_+\gamma\Big|_{{\mathfrak
    h}}=\partial_-\gamma\gamma^{-1}\Big|_{{\mathfrak h}}=0\ .
\label{gco}
}

\section{Magnons by Dressing the Vacuum}
\label{mag}

We shall consider in detail the soliton solutions obtained from a single pair of poles $\{\xi, 1/\xi\}$, with $\xi=r e^{ip/2}$. 
The dressing factor is~\cite{Hollowood:2009tw}
\EQ{
\chi(\lambda)=1+\frac{Q_1}{\lambda-\xi}+\frac{Q_2}{\lambda-1/\xi}\>,
\label{DressFactor}
}
where
\SP{
Q_1&=\frac1\Delta\Big[-\frac{|\xi^2|\beta}{\xi-\xi^*}\BF\BF^\dagger+
\frac{\xi\gamma}
{|\xi|^2-1}\newf_0\theta\BF\BF^\dagger\Big]\ ,\\
Q_2&=\frac1\Delta\Big[\frac{\beta}{\xi-\xi^*}\newf_0\theta
\BF\BF^\dagger\theta \newf_0^\dagger
-\frac{\xi^*\gamma}
{|\xi|^2-1}\BF\BF^\dagger\theta \newf_0^\dagger\Big]\ ,
\label{Qeqs}
}
and we have defined the real numbers
\EQ{
\beta=\BF^\dagger\BF\ ,\qquad\gamma=\BF^\dagger \newf_0\theta\BF\ ,\qquad\Delta=\frac{|\xi|^2\gamma^2}{(|\xi|^2-1)^2}
-\frac{|\xi|^2\beta^2}{(\xi-\xi^*)^2}\ .}
Here,
\EQ{
\BF = \Psi_0(\xi^\ast)\>\Bvarpi
}
where $\Bvarpi$ is a complex $n+1$~dimensional vector.
Then, the magnon solution is given by
$\newf=\Psi(0)=\chi(0)\newf_0$, which corresponds to the projective
coordinates~\cite{Hollowood:2009tw} (see also~\cite{Sasaki:1984tp})
\EQ{
\BZ=\big(\tilde\alpha+\theta\BF\BF^\dagger \theta\big)\BZ_0\>,
\label{Magnon}
}
with
\EQ{
\tilde\alpha=-\frac{\xi\beta}{\xi-\xi^*}-\frac\gamma{|\xi|^2-1}\>.
}

In the following, it will be useful to introduce the notation
\EQ{
f(\xi^\ast)=-i\Big(\frac{x_+}{1+\xi^\ast}+ \frac{x_-}{1-\xi^\ast}\Big) =2F(t,x) - i G(t,x)\>,
\label{Notation}
}
where
\SP{
&
F(t,x) =\frac{1}{2} x' \cos\alpha=\frac{(1+r^2) r \sin\frac{p}{2}}{(1-r^2)^2 + 4r^2 \sin^2\frac{p}{2}}\> x -\frac{r^2 \sin p}{(1-r^2)^2 + 4r^2 \sin^2\frac{p}{2}}\>t
\>,\\[5pt]
&
G(t,x)=t -t'  \sin\alpha=-\frac{2(1-r^2)r \cos\frac{p}{2}}{(1-r^2)^2 + 4r^2 \sin^2\frac{p}{2}}\>x + \frac{2(1-r^2\cos p)}{(1-r^2)^2 + 4r^2 \sin^2\frac{p}{2}}\>t \>.
\label{FandG}
}
The rapidity $\vartheta$ and the parameter $\alpha$ are determined in terms of $r$ and $p$ by
\EQ{
\tanh\vartheta= \frac{2r}{1+r^2}\cos\frac p2\>, \qquad
\cot\alpha= \frac{2r}{1-r^2} \sin\frac{p}{2}\>,
\label{rap}
}
and the Lorentz boosted coordinates $t'$ and $x'$ are
\EQ{
t'=t\cosh\vartheta -x \sinh\vartheta\>, \qquad
x'=x\cosh\vartheta -t \sinh\vartheta
\>.
}

Looking at~\eqref{Qeqs}, it is easy to see that the solutions 
corresponding to $\Bvarpi$ and  
$\Bvarpi\to\lambda \Bvarpi$, $\lambda\in{\mathbb C}$, are equivalent
solutions. The components of $\Bvarpi$ 
represent a set of collective coordinates
for the magnons whose interpretation will be clarified in
Section~\ref{Mother}. In particular, some of the components of
$\Bvarpi$ fix the position of the centre of the
magnons/solitons. Their localized nature arises because $\xi$ has a
imaginary part and thus $\Psi_0(\xi^\ast)$ has an exponential
dependence on $x$. Since $\Lambda$ is anti-hermitian, the relevant
dependence is~\cite{Hollowood:2009tw} 
\EQ{
\exp\Big[ i\>\text{Im}\Big(\frac{x_+}{1+\xi^\ast} + \frac{x_-}{1-\xi^\ast}\Big) \Lambda\Big] =\exp\Big[2 i\>F(t,x) \Lambda\Big]\>, 
}
and this leads to exponential fall-off of the energy/charge density away from the centre which is located at the solution of
\EQ{
F(t,x) = F_0\>,
\label{FF0}
}
where $F_0$ is a constant determined by the components of $\Bvarpi$. In particular, this equation shows that the velocity of the magnon is
\EQ{
v = \frac{2r}{1+r^2} \cos\frac{p}{2} = \tanh\vartheta\>.
}
Moreover, since $\Psi_0(\xi^\ast)$ always appear in the combination $\BF= \Psi_0(\xi^\ast)\Bvarpi$,  a constant shift of the solitons in
space and time act on the collective coordinates via\footnote{For the magnon solutions $\newf=\chi(0)\newf_0$, this transformation gives rise to a constant shift of the dressing factor $ \chi(0)$ in space and time, but not of $\newf_0$ which depends only on $t$. In contrast, it is completely equivalent to a constant shift of the solitonic avatars given by~\eqref{xxd} in space and time.}
\EQ{
\Bvarpi\longrightarrow \exp\Big[\Bigl(\frac{\delta
  x_+}{1+\xi^*}+\frac{\delta x_-}
{1-\xi^*}\Bigr)\Lambda\Big]\Bvarpi\>.
\label{com}
}

The solutions provided by~\eqref{Magnon} give rise to different species of magnons. 
In order to find them out, we shall investigate the value of the $SU(n+1)$ charge $\Delta{\cal Q}_L$ by means of~\eqref{Charges}.
Using~\eqref{Notation}, we can write
\EQ{
\Psi_0(\xi^\ast)=\frac{1}{2}\> e^{f(\xi^\ast)} \left(\begin{array}{cc|c} 1&-i&\B0\\ +i&1&\B0\\ \hline \B0&\B0&\B0 \end{array}\right)
+ \left(\begin{array}{cc|c} 0&0&\B0\\ 0&0&\B0\\ \hline \B0&\B0&\boldsymbol{I}\end{array}\right)
+\frac{1}{2}\> e^{-f(\xi^\ast)} \left(\begin{array}{cc|c} 1&+i&\B0\\ -i&1&\B0\\ \hline \B0&\B0&\B0 \end{array}\right)\>,
\label{Explicit}
}
whose asymptotic behaviour can be easily worked out by noticing that 
\EQ{
\lim_{x\rightarrow\pm\sigma\infty} F(t,x)=\pm\infty\qquad
\text{with} \qquad \sigma=\text{sign}\Big(r \sin\frac{p}{2}\Big)\>.
}
Now, if we split the components of the complex $(n+1)$-vector $\Bvarpi$ as
\EQ{
\Bvarpi=\omega_1 \boldsymbol{e}_1 + \omega_2 \boldsymbol{e}_2 + \BOmega\>, \qquad
\BOmega\cdot\boldsymbol{e}_1=\BOmega\cdot\boldsymbol{e}_2=0\>,
}
it is easy to show that
\SP{
&
\beta=\BF^\dagger\BF =\frac{1}{2} e^{4F} |\omega_1 -i\omega_2|^2 + \BOmega^\dagger \BOmega+
\frac{1}{2} e^{-4F} |\omega_1 +i\omega_2|^2 \\[5pt]
&
\gamma=\BF^\dagger\newf_0\theta\BF=\bigl(|\omega_2|^2- |\omega_1|^2\bigr) \cos\bigl(2(G-t)\bigr) \\[5pt]
&\qquad\qquad\qquad\quad
+ \bigl(\omega_1\omega_2^\ast +\omega_1^\ast\omega_2\bigr) \sin\bigl(2(G-t)\bigr) + \BOmega^\dagger \BOmega\>.
\label{betadef}
}
Therefore, we can distinguish three cases.

\subsection{$\boldsymbol{\big|\omega_1\pm i\omega_2\big|\not=0}$.}
\label{original}

In this case,
\SP{
&
\beta \buildrel x\rightarrow\pm\sigma\infty
\over{\hbox to 50pt{\rightarrowfill}}\>
\frac{1}{2} e^{\pm 4F} |\omega_1 \mp i\omega_2|^2+ \cdots
\>, \qquad
\gamma \buildrel x\rightarrow\pm\sigma\infty
\over{\hbox to 50pt{\rightarrowfill}}\> \text{finite}
\\[5pt]
&
\Longrightarrow\Delta
\buildrel x\rightarrow\pm\sigma\infty
\over{\hbox to 50pt{\rightarrowfill}}\>
-\frac{|\xi|^2}{(\xi-\xi^\ast)^2}\>
\frac{1}{4} e^{\pm 8F} |\omega_1 \mp i\omega_2|^2+ \cdots\>,
}
and
\SP{
&
\BF\BF^\dagger \buildrel x\rightarrow\pm\sigma\infty
\over{\hbox to 50pt{\rightarrowfill}}\>
\frac{1}{4} e^{\pm 4F} |\omega_1 \mp i\omega_2|^2\>\left(\begin{array}{cc|c}1 & \mp i & \B0\\ 
\pm i &1 & \B0 \\ \hline
\B0 & \B0  &
\B0\end{array}\right) + \cdots\\[5pt]
&
\newf_0\theta\BF\BF^\dagger\theta\newf_0^\dagger \buildrel x\rightarrow\pm\sigma\infty
\over{\hbox to 50pt{\rightarrowfill}}\>
\frac{1}{4} e^{\pm 4F} |\omega_1 \mp i\omega_2|^2\>\left(\begin{array}{cc|c}1 & \pm i & \B0\\ 
\mp i &1 & \B0 \\ \hline
\B0 & \B0  &
\B0\end{array}\right) + \cdots\>.
}
Therefore,
\SP{
&
Q_1\buildrel x\rightarrow\pm\sigma\infty
\over{\hbox to 45pt{\rightarrowfill}}\>
\frac{1}{2}(\xi-\xi^\ast)\>\left(\begin{array}{cc|c}1 & \mp i & \B0\\ 
\pm i &1 & \B0 \\ \hline
\B0 & \B0  &
\B0\end{array}\right)\>, \qquad
Q_2\buildrel x\rightarrow\pm\sigma\infty
\over{\hbox to 45pt{\rightarrowfill}}\>
-\frac{1}{2}\frac{\xi-\xi^\ast}{|\xi|^2}\>\left(\begin{array}{cc|c}1 & \pm i & \B0
\\
\mp i &1 & \B0 \\ \hline
\B0 & \B0  &
\B0\end{array}\right)
}
which, using~\eqref{Charges}, give rise to the charge
\EQ{
\Delta{\cal Q}_L= \Bigl(Q_1+Q_2\Bigr)\Big|_{-\infty}^{+\infty} =\sigma\left[ i(\xi-\xi^\ast) \Lambda + i\frac{\xi-\xi^\ast}{|\xi|^2}\Lambda\right]
=-2\>\frac{r^2+1}{|r|}\big|\sin\frac{p}{2}\big|\Lambda\>.
\label{MagnonCharge1}
}

This case produces the magnon solution constructed in~\cite{Hollowood:2009tw} (see also~\cite{Kalousios:2009mp,Suzuki:2009sc}). 
Notice that the transformation~\eqref{com}, which corresponds to a
constant shift in space and time, is equivalent to
\EQ{
\omega_1 \pm i\omega_2 \longrightarrow e^{\mp \delta f} \big(\omega_1 \pm i\omega_2\big)
\label{TransA}
}
where (see~\eqref{Notation})
\EQ{
\delta f=-i\Big(\frac{\delta x_+}{1+\xi^\ast}+ \frac{\delta x_-}{1-\xi^\ast}\Big) =2\delta F - i \delta G\>.
\label{DeltaF}
}
Then,
\EQ{
\Big| \frac{\omega_1-i\omega_2}{\omega_1+i\omega_2}\Big| \longrightarrow e^{4 \delta F} \Big| \frac{\omega_1-i\omega_2}{\omega_1+i\omega_2}\Big| 
}
and, as explained in the paragraph around~\eqref{FF0}, the centre of this magnon is located at the solution of
\EQ{
F(t,x)= \frac{1}{4}\log \Big| \frac{\omega_1-i\omega_2}{\omega_1+i\omega_2}\Big|\>,
\label{CentreOld}
}
which clarifies the meaning of the collective coordinates $\omega_1$ and $\omega_2$. Then, up to a shift of the soliton in space and time, we can always set $\omega_2=0$ and, using the invariance under complex re-scalings $\Bvarpi\rightarrow\lambda\Bvarpi$, we can also set $\omega_1=i$, so that
\EQ{
\Bvarpi=i\boldsymbol{e}_1+\BOmega\>,
}
which is precisely the normalization used in~\cite{Hollowood:2009tw}.

\subsection{$\boldsymbol{\big|\omega_1+ i\omega_2\big|=0}$.}
\label{NewSolSec}

In this case, $\omega_2=+i\omega_1$ and thus
\SP{
&\beta= 2|\omega_1|^2 e^{+ 4F} + \BOmega^\dagger\BOmega\>, \qquad
\gamma=\BOmega^\dagger\BOmega\\[5pt]
&
\Longrightarrow\Delta
\buildrel x\rightarrow+\sigma\infty
\over{\hbox to 50pt{\rightarrowfill}}\>-\frac{|\xi|^2}{(\xi-\xi^\ast)^2}\>
4 e^{+ 8F}+ \cdots\\[5pt]
&
\qquad\;
\Delta
\buildrel x\rightarrow-\sigma\infty
\over{\hbox to 50pt{\rightarrowfill}}\>\Bigl[\frac{|\xi|^2}{(|\xi|^2-1)^2}-\frac{|\xi|^2}{(\xi-\xi^\ast)^2}\Bigr]\> \bigl(\BOmega^\dagger\BOmega\bigr)^2+ \cdots\>,
}
together with
\SP{
&
\BF\BF^\dagger \buildrel x\rightarrow+\sigma\infty
\over{\hbox to 50pt{\rightarrowfill}}\>
|\omega_1|^2e^{+ 4F}\>\left(\begin{array}{cc|c}1 & - i & \B0\\ 
+ i &1 & \B0 \\ \hline
\B0 & \B0  &
\B0\end{array}\right) + \cdots\\[5pt]
&
\newf_0\theta\BF\BF^\dagger\theta\newf_0^\dagger \buildrel x\rightarrow+\sigma\infty
\over{\hbox to 50pt{\rightarrowfill}}\>
|\omega_1|^2 e^{+ 4F}\>\left(\begin{array}{cc|c}1 & + i & \B0\\ 
- i &1 & \B0 \\ \hline
\B0 & \B0  &
\B0\end{array}\right) + \cdots\\[5pt]
&
\BF\BF^\dagger\>, \;\newf_0\theta\BF\BF^\dagger\theta\newf_0^\dagger\>, \;\newf_0\theta\BF\BF^\dagger\>, \;\BF\BF^\dagger\theta\newf_0^\dagger
\buildrel x\rightarrow-\sigma\infty
\over{\hbox to 50pt{\rightarrowfill}}\>\left(\begin{array}{c|c}\B0 & \B0\\ 
\hline
\B0 & \BOmega\BOmega^\dagger\end{array}\right) + \cdots
}
This leads to
\SP{
&
Q_1\buildrel x\rightarrow+\sigma\infty
\over{\hbox to 50pt{\rightarrowfill}}\>
\frac{1}{2}(\xi-\xi^\ast)\>\left(\begin{array}{cc|c}1 & - i & \B0\\ 
+ i &1 & \B0 \\ \hline
\B0 & \B0  &
\B0\end{array}\right)\>, \qquad
Q_2\buildrel x\rightarrow+\sigma\infty
\over{\hbox to 50pt{\rightarrowfill}}\>
-\frac{1}{2}\frac{\xi-\xi^\ast}{|\xi|^2}\>\left(\begin{array}{cc|c}1 & + i & \B0
\\
- i &1 & \B0 \\ \hline
\B0 & \B0  &
\B0\end{array}\right)\>,\\[5pt]
&Q_1+Q_2 \buildrel x\rightarrow-\sigma\infty
\over{\hbox to 50pt{\rightarrowfill}}\> \frac{(\xi-\xi^\ast)(|\xi|^2-1)}{|\xi|^2}
\>\left(\begin{array}{c|c}\B0 & \B0
\\\hline
\B0 &{\BOmega\BOmega^\dagger}/{\BOmega^\dagger\BOmega}
\end{array}\right)
}
and, using~\eqref{Charges}, to the charge
\EQ{
\Delta{\cal Q}_L= \Bigl(Q_1+Q_2\Big)\Big|_{-\infty}^{+\infty} =-\>\frac{1+r^2}{|r|}\big|\sin\frac{p}{2}\big|\Lambda \>-\> \frac{1-r^2}{|r|}\big|\sin\frac{p}{2}\big| \>\left(\begin{array}{c|c}i\boldsymbol{1} & \B0
\\ \hline
\B0 &{-2i\BOmega\BOmega^\dagger}/{\BOmega^\dagger\BOmega} 
\end{array}\right)\>.
\label{NewCharge}
}
It corresponds to a new dyonic solution specified by the projective coordinates
\SP{
\BZ&= \Big[-\big( 2|\omega_1|^2e^{4F} + \BOmega^\dagger \BOmega\bigr)\frac{\xi}{\xi-\xi^\ast}- \frac{\BOmega^\dagger \BOmega}{|\xi|^2-1}\Big] \big(\cos t \>\boldsymbol{e}_1 -\sin t\> \boldsymbol{e}_2\bigr)\\[5pt]
& + |\omega_1|^2e^{4F} e^{-it} \big(\boldsymbol{e}_1-i \boldsymbol{e}_2\bigr)
- \omega_1^\ast e^{2F} e^{i(G-t)}\BOmega\>,
\label{NewSol}
}
which is apparently singular when $|\xi|=1$. However, a regular solution in this limit can be constructed by imposing the additional condition that
\EQ{
\BOmega^\dagger \BOmega=0\; \Rightarrow \BOmega=0\>.
}
This particular limit produces precisely the embedding of the
Hofman-Maldacena $S^2$ magnon in $\CP^1$ considered
in~\cite{Gaiotto:2008cg}.  
In fact,~\eqref{NewSol} is the dyonic generalization of the $\CP^1$ magnon
whose existence was conjectured in~\cite{Abbott:2009um} where only the
explicit form of a solution with $p=\pi$ was provided. In
appendix~\ref{AASappendix}, we show that the solution reported
in~\cite{Abbott:2009um} is recovered for the particular choice of
parameters  
\EQ{
p=\pi\>, \qquad
\BOmega^\dagger \BOmega=2\>\frac{r^2-1}{r^2+1}|\omega_1|^2\>.
}

In this case, the transformation~\eqref{com}, which corresponds to a
constant shift in space and time, is equivalent to
\EQ{
\omega_1 \longrightarrow e^{+\delta f} \omega_1\;\Rightarrow \; |\omega_1| \longrightarrow e^{2\delta F}|\omega_1|
\label{TransA}
}
where $\delta f$ is defined in~\eqref{DeltaF}.
Then, according to the discussion around~\eqref{FF0},  and taking into account that the solution depends on $\omega_1$ and $\BOmega$ only up to complex re-scalings $\omega_1\to\lambda\omega_1$ and $\BOmega\to\lambda\BOmega$, the centre of this magnon is located at the solution of
\EQ{
F(t,x)= \frac{1}{2}\log \frac{|\omega_1|}{|\BOmega|}\>.
\label{CentreNew}
}
Therefore, up to a shift of the soliton in space and time, we can always set $|\BOmega|=|\omega_1|$ and, using the invariance under complex re-scalings $\Bvarpi\rightarrow\lambda\Bvarpi$, we can also set $\omega_1=1$, so that
\EQ{
\Bvarpi=\boldsymbol{e}_1+ i\boldsymbol{e}_2+\BOmega\>,
}
with $|\BOmega|=1$.

Eq.~\eqref{NewSol} manifests the fact that the new solution takes
values in the $\CP^2\subset \CP^n$ subspace picked out by the three
mutually orthogonal vectors
$\{\boldsymbol{e}_1,\boldsymbol{e}_2,\BOmega\}$. According
to~\eqref{NewCharge}, its charge has two independent
components. Namely, 
\EQ{
\Delta{\cal Q}_L = J_\Lambda \Lambda + J_h h_{\BOmega}\>,\qquad J_\Lambda=-\frac{1+r^2}{|r|}\big|\sin\frac{p}{2}\big|\>, \qquad
J_h= -\frac{1-r^2}{|r|}\big|\sin\frac{p}{2}\big|\>,
\label{NewCharge2}
}
where 
\EQ{
h_\BOmega=\left(\begin{array}{c|c}i\boldsymbol{1} & \B0
\\ \hline
\B0 &{-2i\BOmega\BOmega^\dagger}/{\BOmega^\dagger\BOmega} 
\end{array}\right)
\label{defh}
}
is one of the generators of $H=U(n-1)$, which is the subgroup of elements of $G=U(n)$ that commute with $\Lambda$.
These charges satisfy the relation
\EQ{
-J_\Lambda = \sqrt{J_h^2 + 4\sin^2\frac{p}{2}}\>.
\label{Dispersion}
}
In the AdS/CFT context~\cite{Hofman:2006xt,DyonicGM}, the components $J_\Lambda$ and $J_h$ are identified, up to scaling, with $\Delta-\frac{1}{2}J$ and $Q$, respectively, where $\Delta$ is the scaling dimension of the associated operator in the CFT, and $J$ and $Q$ are two conserved $U(1)$ $R$-charges:\footnote{To be specific, we use the same normalization as~\cite{Abbott:2009um}.}
\EQ{
\Delta -\frac{1}{2}J = -\sqrt{\frac{\lambda}{2}}\> J_\Lambda\>, \qquad
\frac{1}{2}Q = \sqrt{\frac{\lambda}{2}}\> J_h\>,
}
where $\lambda$ is the 't~Hooft coupling. Then,~\eqref{Dispersion} becomes the celebrated dispersion relation
\EQ{
\Delta -\frac{1}{2}J = \sqrt{\frac{1}{4}Q^2 + 2\lambda\sin^2\frac{p}{2}}\>.
}

Eq.~\eqref{NewCharge2} shows that all the inequivalent magnons of
this type are obtained with $r>0$ and $\sin\frac{p}{2}>0$, and that the transformation
$r\rightarrow 1/r$ corresponds to $J_h\rightarrow -J_h$.

\subsection{$\boldsymbol{\big|\omega_1- i\omega_2\big|=0}$.}
\label{NewSolSec2}

This is equivalent to $\omega_2=-i\omega_1$. Again, using the invariance under complex re-scalings $\Bvarpi\rightarrow\lambda\Bvarpi$, we can set $\omega_1=1$, so that
\EQ{
\Bvarpi=\boldsymbol{e}_1 -i\boldsymbol{e}_2 +\BOmega\>.
}
However, this case does not give rise to new magnon solutions. Let us write
\EQ{
\widehat  \Bvarpi = \left(\begin{array}{c}\omega_1\\ -i\omega_1\\ 
\BOmega\end{array}\right) = -\theta\left(\begin{array}{c}\omega_1\\ +i\omega_1\\ 
-\BOmega\end{array}\right)\equiv -\theta \Bvarpi\>,
}
where $\Bvarpi$ produces~\eqref{NewSol} with $\BOmega\to-\BOmega$. Then, using~\eqref{CondPsi},
\EQ{
\BF[\widehat \Bvarpi,\xi] = \Psi_0(\xi^\ast) \widehat  \Bvarpi = -\theta \newf_0^{-1} \Psi_0(1/\xi^\ast) \Bvarpi = -\theta \newf_0^{-1} \BF[\Bvarpi,1/\xi]\>,
}
where we have explicitly indicated the dependence of $\BF$ on $\Bvarpi$ and $\xi$. This implies
\EQ{
\beta[\widehat \Bvarpi,\xi]=\beta[\Bvarpi,1/\xi]\>, \qquad
\gamma[\widehat \Bvarpi,\xi]=\gamma[\Bvarpi,1/\xi]\>,\qquad
\Delta[\widehat \Bvarpi,\xi]=\Delta[\Bvarpi,1/\xi]
}
and, finally,
\EQ{
Q_1[\widehat \Bvarpi,\xi] = Q_2[\Bvarpi,1/\xi]\>.
}
Therefore, the dressing factor\eqref{DressFactor} satisfies
\EQ{
\chi[\lambda;\> \widehat \Bvarpi,\xi] =\chi[\lambda;\> \Bvarpi,1/\xi]\>,
\label{Equivalence}
}
which shows that $\widehat \Bvarpi=-\theta\Bvarpi$ and $\xi$ gives rise to the same solution as $\Bvarpi$ and $1/\xi$,
and confirms that taking $\omega_2=- i\omega_1$ also leads to solutions of the form~\eqref{NewSol}. It is worth noticing that $f(1/\xi^\ast)= -f(\xi^\ast) -2it$, which means that\EQ{
\xi\to 1/\xi\; \Rightarrow \;
F(t,x)\longrightarrow -F(t,x)\>, \qquad G(t,x)\longrightarrow -G(t,x)+2t\>.
}

\section{The Solitonic Avatars}
\label{avatar}

In the present case, it is sufficient to
consider the case of $\CP^2$ because the solutions in this case
 can then be simply
embedded in $\CP^n$, $n>2$, in the obvious way. 
Introducing the parameterization in 
  \cite{Hollowood:2009tw} 
\EQ{
\gamma=e^{a_L h}\MAT{1&0&0\\ 0&\cos\theta e^{i\varphi}&\sin\theta\\
0&-\sin\theta &\cos\theta e^{-i\varphi}}
e^{-a_R h}\ ,
\label{ksa}
}
where $h=i\,\text{diag}(1,1,-2)$ is the generator of $\mathfrak h$,
the Lie algebra of $H=U(n-1)$, the dyon solution~\eqref{xxd} in the rest frame $(p=\pi$) is\footnote{In the
  following, we have used the symmetries to shift
$x$, $\psi$ and $\tilde\psi$ by appropriate constants.}
\SP{
\tilde\psi&=\frac43\frac{1-r^2}{r^2+1}t\ ,\\
\psi&=-4\tan^{-1}\left(\frac1r\tanh\frac{4rx}{r^2+1}\right)\ ,\\
\varphi&=-\frac23\tan^{-1}\left(\frac1r\right)
-\tan^{-1}\left(\frac{1-3r^2}{r^3-3r}\tanh\frac{4rx}{r^2+1}
\right)+\tan^{-1}\left(\frac1r\tanh\frac{4rx}{r^2+1}\right)\ ,\\
\sin^2\theta&=\frac{16r^2(r^2-1)}
{(r^2+1)^2\left(2r^2+(r^2+1)\sinh^2\frac{4rx}{r^2+1}\right)}\ ,
\label{sol}
}
where
\EQ{
\psi=2(a_L-a_R)\ ,\qquad\tilde\psi=2(a_L+a_R)\ .
}
The field $\psi(x,t)$ has a kink-like behaviour with
\EQ{
\Delta\psi=\psi(\infty,t)-\psi(-\infty,t)=-8\tan^{-1}\big(r^{-1}\big)=
-2\pi-4\alpha\ ,
}
where $\alpha$ is defined by \eqref{rap} for $p=\pi$ (vanishing
rapidity), so $\cot\alpha=2r/(1-r^2)$. Notice that the $S^1$-valued
field $\tilde\psi(x,t)$ has constant angular velocity
$\tfrac43(1-r^2)/(r^2+1)=\tfrac43\sin\alpha$. In this sense the
solution is metaphorically the dyon solution of four-dimensional
gauge theories where the angular variable $\tilde\psi$ 
is the $U(1)$ charge angle of the magnetic monopole.

As described in \cite{Hollowood:2009tw,Miramontes:2008wt}, 
the SSSG system can be written as the equations-of-motion of a
Lagrangian field theory. For the $\CP^2$ theory there are two
inequivalent ways of doing this corresponding to ``vector gauging''
and ``axial gauging'' which are related by a target-space T-duality symmetry~\cite{Miramontes:2004dr}. (For $\CP^n$, $n>2$, axial gauging is not possible.)
For vector gauging, the field $\tilde\psi(x,t)$ is a gauge
degree-of-freedom and is consequently ``gauged away''
leaving a theory with physical fields $(\varphi,\psi,\theta)$ and a
Lagrangian density
\EQ{
\LAG=\partial_\mu\theta\partial^\mu\theta+\frac14\partial_\mu\psi
\partial^\mu\psi+\cot^2\theta\partial_\mu(\psi+\varphi)\partial^\mu(
\psi+\varphi)+2  \cos\theta\cos\varphi\ .
}
The $H=U(1)$ symmetry $\psi\to\psi+a$ is broken by the vacuum
configuration $\theta=0$, $\varphi=0$ and $\psi$ arbitrary, and
consequently the soliton \eqref{sol} carries
topological, or kink, charge 
\EQ{
Q^T=\frac14\Delta\psi\,h=\Big(-\frac \pi2-\alpha\Big)h\ .
}

On the contrary, for the axial-gauged theory, the field 
$\psi(x,t)$ is a gauge degree-of-freedom that is gauged away leaving a
theory with physical fields $(\varphi,\tilde\psi,\theta)$ and a 
Lagrangian density
\SP{
\LAG=\partial_\mu\theta\partial^\mu\theta+\frac1{1+4\cot^2\theta}&
\Big(\frac94\partial_\mu\tilde\psi
\partial^\mu\tilde\psi+\cot^2\theta\partial_\mu\varphi\partial^\mu
\varphi\\ &
-6\cot^2\theta\epsilon^{\mu\nu}\partial_\mu\tilde\psi\partial_\nu
\varphi\Big)+2  \cos\theta\cos\varphi\ ,
\label{axiallag}
}
In the resulting theory, the $H=U(1)$ group survives as a
genuine symmetry $\tilde\psi\to\tilde\psi+a$ 
(corresponding to vector transformations) 
with an associated Noether current. The vacuum is invariant under the
symmetry and remains unbroken. 
The quantity $\psi(x,t)$ still
plays a r\^ole since it determines the associated Noether current
\EQ{
J^\mu=\epsilon^{\mu\nu}({\cal A}_\nu+\tfrac14\partial_\nu\psi\,h)\ ,
} 
In this case the gauge field vanishes and so, as expected, 
the Noether charge is the same as the topological charge of the 
vector-gauged theory:
\EQ{
Q^N=\int dx \,J_0=\frac14\Delta\psi h
=\Big(-\frac \pi2-\alpha\Big)h\ .
}

The masses of the dyon follows from the
additivity of the mass of the---now realized to be composite (see Section~\ref{Mother})---soliton
calculated in \cite{Hollowood:2009tw}, giving
\EQ{
M_\text{dyon}=\frac{4r}{r^2+1}=2\cos\alpha\ .
\label{mdy}
}
It is interesting to note that the dyon is a generalization of the
soliton (or dyon) of the complex sine-Gordon theory
\cite{Dorey:1994mg}. It would be interesting to follow the approach
of \cite{Dorey:1994mg} to quantize the $\CP^2$ model (see also~\cite{CastroAlvaredo:2000kq}).

\section{Constructing all the known $\boldsymbol{\CP^n}$ Magnons}
\label{Mother}

Since these new objects carry general charges
we refer to them as ``$\CP^n$ dyons''. They are labelled by the data
$\big(r,p,\BOmega\big)$, corresponding to dressing data
$\Bvarpi=\ee_1+i\ee_2+\BOmega$ and $\xi=re^{ip/2}$,
which determines their charges and rapidity
(the latter via \eqref{rap}). The overall magnitude and phase of
$\BOmega$ determines the position and $U(1)$ ``charge angle'' of the
dyon, respectively, leaving the equivalence class
$\CP^{n-2}=\{\BOmega\sim \lambda\BOmega,
\lambda\in\C\not=0\}$ to specify its non-abelian orientation.\footnote{In
  more detail we  can think of $\BOmega$ as a real $2n-2$ vector. 
The moduli space 
${\mathbb R}^{2n-2}\simeq {\mathbb R}^+\times S^{2n-3}$, for which the
radius factor determines the dyon's position. The $S^{2n-3}$ can be
viewed as a Hopf fibration of $S^1$ over a $\CP^{n-2}$ and the $S^1$
angle is associated to the $U(1)$ factor of the group $H=U(n-1)$ (the
centralizer of $\Lambda$ in $G=U(n)$), this
is the ``charge angle'' of the dyon which is rotating with constant
angular velocity, while the
$\CP^{n-2}\simeq SU(n-1)/U(n-2)$ 
factor describes the orientation of the dyon in the non-abelian
subgroup $SU(n-1)\subset H$.} These new dyonic magnons are 
the most elementary type of solution because all the other known
solutions are either composites of them, or are obtained by taking
special limits, as we now explain seriatim:

(i) The original $\CP^n$ magnon of \cite{Hollowood:2009tw} 
corresponds to a configuration of two of the new dyons of the form
\EQ{
\big(r,p,\tfrac12\BOmega\big) + \big(1/r,-p,-\tfrac12\BOmega\big)\>.
}
In order to see this, an inspection of
Eqs.~\eqref{MagnonCharge1} and~\eqref{NewCharge} illustrates precisely that
the magnon solution originally constructed in~\cite{Hollowood:2009tw}
(and in Section~\ref{original}) is to  
be thought of as a composite of two new dyonic magnons.
In order to make this explicit, consider~\eqref{Explicit}, 
which leads to
\EQ{
\BF =\Psi_0(\xi^\ast)\Bvarpi= \frac{\omega_1-i\omega_2}{2}\> e^{f(\xi^\ast)} \left(\begin{array}{c}1\\ +i\\ \hline \B0\end{array}\right) + \left(\begin{array}{c}0\\ 0\\ \hline \BOmega\end{array}\right) 
+\frac{\omega_1+i\omega_2}{2}\> e^{-f(\xi^\ast)} \left(\begin{array}{c}1\\ -i\\ \hline \B0\end{array}\right)\>.
}
Then, since the solutions are constructed in terms of $\BF$, we can
indeed interpret this one as a superposition of two more basic
constituents which are simply two dyonic magnons like those
constructed in  \ref{NewSolSec} associated to $(\xi, \tfrac12\BOmega)$ and
$(1/\xi, -\tfrac12\BOmega)$. Using~\eqref{CentreNew}, they are mutually at rest at space-time
positions determined by\footnote{Recall that $F\to-F$ under
  $\xi\to1/\xi$.} 
\EQ{
F_1(t,x_1)= \frac{1}{2} \log\Big(\frac{|\omega_1-i\omega_2|}{|\BOmega|}\Big)\>,\qquad
-F_2(t,x_2)= \frac{1}{2}
\log\Big(\frac{|\omega_1+i\omega_2|}{|\BOmega|}\Big)\ .
}
Then, the centre of the composite soliton is at
\EQ{
F(t,x)=\frac{1}{2}\big(F_1(t,x_1)+F_2(t,x_2) \big)= \frac{1}{4} \log\Big|\frac{\omega_1-i\omega_2}{\omega_1+i\omega_2}\Big|\>,
}
in agreement with~\eqref{CentreOld}. Moreover, the distance
between the two constituents is
\EQ{
\Delta x=x_1-x_2=\sec\alpha\> \sech\vartheta\> \log\Big(\frac{|\omega_1^2+\omega_2^2|}
{\BOmega^\dagger\BOmega}\Big)\ .
}
The charges of the two constituents 
in the direction $h_\BOmega$ cancel to leave 
\eqref{MagnonCharge1}.

(ii) The embedding of the Hofman-Maldacena magnon in $\CP^1$ is
recovered as the $r\rightarrow 1$, $\BOmega\rightarrow \B0$ limit of the
new dyon. 

(iii) The embedding of the Hofman-Maldacena magnon in $\RP^2$ is
recovered as the $r\rightarrow 1$, $|\BOmega|\to 1$ limit of the magnon in (i).

(iv) The final type of solution is the embedding of Dorey's
dyonic magnon in $\RP^3$. This solution turns out to be a composite of
two of the elementary dyons of the form
\EQ{
\big(r,p,\BOmega^{(1)}-i\BOmega^{(2)}\big) + 
\big(1/r,p,\BOmega^{(1)}+i\BOmega^{(2)}\big)\>,
\label{assign}
}
where $\BOmega^{(i)}$, $i=1,2$ are two mutually orthogonal unit vectors,
whose charge is
\EQ{
\Delta{\cal Q}_L=-\>2\frac{1+r^2}{|r|}\big|\sin\frac{p}{2}\big|\Lambda
- 2\frac{(1-r^2)}{|r|}\big|\sin\frac{p}{2}\big| \>
\left(\begin{array}{cc|c} 0 & 0&\B0^T
\\ 0&0&\B0^T\\ \hline \B0 &\B0&
\BOmega^{(1)}\BOmega^{(2)T}-\BOmega^{(2)}\BOmega^{(1)T}\end{array}\right)\>.
}
Actually, the fact that Dorey's magnon is a composite object could
have been guessed by considering the solitonic avatars. The new
elementary dyon has a mass, eq.~\eqref{mdy},
$M_\text{dyon}=2\cos\alpha$, whereas the avatar of Dorey's magnon has a
mass $M_\text{Dorey}=4\cos\alpha$ \cite{Hollowood:2009tw}.

From the point-of-view of the dressing transformation Dorey's dyonic
magnon corresponds to a solution with 4 poles at
$\lambda_i=(\xi,1/\xi,1/\xi^*,\xi^*)$ along with 
$\mu_i=(\xi^*,1/\xi^*,1/\xi,\xi)$. 
The first elementary dyon
corresponds to the pair $\lambda_1=\xi$ and $\lambda_2=1/\xi$ 
and the second to $\lambda_3=1/\xi^*$ and $\lambda_4=\xi^*$
and the associated vectors $\Bvarpi_i$, $i=1,\ldots,4$, have the
pair-wise constraints
\EQ{
\Bvarpi_2=\theta\Bvarpi_1\ ,\qquad \Bvarpi_4=\theta\Bvarpi_3\ .
}
However, because $\mu_1=\lambda_4$ and $\mu_2=\lambda_3^*$ there are
the additional constraints that require
\EQ{
\Bvarpi_1^\dagger \Bvarpi_4=0\ ,\qquad\Bvarpi_2^\dagger \Bvarpi_3=0\ .
\label{addcon}
}
The individual elementary dyons therefore have the dressing data\footnote{As
  previously, $\BOmega_1$ and $\BOmega_2$ are orthogonal to $\ee_1$
  and $\ee_2$.} 
\EQ{
\Bvarpi_1=\ee_1+i\ee_2+\BOmega_1\ ,\qquad
\Bvarpi_2=-\ee_1+i\ee_2+\BOmega_1\ ,
}
and
\EQ{
\Bvarpi_3=\nu\big(-\ee_1-i\ee_2+\BOmega_2\big)\ ,\qquad
\Bvarpi_4=\nu(\ee_1-i\ee_2+\BOmega_2\big)\ ,
}
where we have used the overall-scaling symmetry to fix the scaling of
$\Bvarpi_{1,2}$. Then the additional constraints \eqref{addcon} require 
\EQ{
\BOmega_1^\dagger\cdot\BOmega_2=0\ .
\label{Orthog}
}
Dorey's dyon is then obtained as
the special case when $\nu=1$ and with the choice 
\EQ{
\BOmega_1=\BOmega^{(1)}-i\BOmega^{(2)}\ ,\qquad
\BOmega_2=\BOmega^{(1)}+i\BOmega^{(2)}\ ,
\label{cvv}
}
for two orthogonal real vectors $\BOmega^{(1,2)}$ with
$|\BOmega^{(1)}|=|\BOmega^{(2)}|$. Notice this solves the constraint
above. One can view setting $\nu=1$ as fixing the
relative position of the two elementary dyons in $x$ and $t$.
In addition, by choosing a suitable overall origin in
$x$ and $t$ the two vectors $\BOmega^{(1,2)}$
can be chosen to have unit length leading to the 
assignments in \eqref{assign}.

It is important to emphasize
that in the context of $\CP^n$, Dorey's dyon is a composite
solution since $\nu$ can be thought of as determining the relative
position of the two constituent dyons in $x$ and $t$. 
However, the composite solution is also a solution of the sigma model
with target space $S^n$. The dressing transformation in this case is
identical up to the fact that one has to impose an additional reality
condition on the solutions. This additional constraint leads to the
conditions on the dressing data:
\EQ{
\Bvarpi_4=\Bvarpi_1^*\ ,\qquad \Bvarpi_3=\Bvarpi_2^*\ ,
}
which enforces the choice $\nu=1$ and \eqref{cvv}. 
In other words, the additional reality
constraint locks the two elementary dyons together to form
Dorey's---now elementary---dyon. 
This is as it should be since the new elementary dyon is not by
itself a solution in the $S^n$ theory.

\acknowledgments

\noindent TJH would like to acknowledge the support of STFC grant
ST/G000506/1. The work of JLM was partially  supported by MICINN (Spain) 
and FEDER (FPA2008-01838 and 
FPA2008-01177), by Xunta de Galicia (Consejer\'\i a de 
Educaci\'on and PGIDIT06PXIB296182PR), and by the 
Spanish Consolider-Ingenio 2010
Programme CPAN (CSD2007-00042). 
JLM thanks Diego Bombardelli for helpful discussions.

\startappendix

\Appendix{The Generalized $\boldsymbol{\CP^2}$ Dyon of~\cite{Abbott:2009um}} \label{AASappendix}

In this appendix we shall recover the $\CP^2$ solution constructed by Abbott, Aniceto and Sax (AAS) in~\cite{Abbott:2009um} from~\eqref{NewSol}.

Consider~\eqref{NewSol} for $p=\pi$:
\SP{
\BZ&= \Big[-\frac{1}{2}\>\big( 2\> |\omega_1|^2 e^{\frac{4rx}{r^2+1}}+ \BOmega^\dagger \BOmega\bigr)- \frac{\BOmega^\dagger \BOmega}{r^2-1}\Big] \big(\cos t \>\boldsymbol{e}_1 -\sin t\> \boldsymbol{e}_2\bigr)\\[5pt]
& + |\omega_1|^2 e^{\frac{4rx}{r^2+1}}\> e^{-it} \big(\boldsymbol{e}_1-i \boldsymbol{e}_2\bigr)
- e^{\frac{2rx}{r^2+1}}\>  e^{-i\frac{r^2-1}{r^2+1}t}\> \omega_1^\ast \BOmega\>.
}
In order to simplify the comparison with the solution proposed in~\cite{Abbott:2009um}, we perform a $U(n+1)$ rotation 
\EQ{
\BZ\longrightarrow U\BZ\>, \qquad
U=\left(\begin{array}{cc|c}
\frac{1}{\sqrt{2}}& -\frac{i}{\sqrt{2}}~ & \B0\\
\frac{1}{\sqrt{2}}&+\frac{i}{\sqrt{2}}~&\B0\\[4pt] \hline \B0&\B0&{\bf 1}\end{array}\right)
}
so that
\SP{
\widetilde \BZ =U\BZ =&
-e^{it}\; \frac{2(r^2-1) |\omega_1|^2 e^{\frac{4rx}{r^2+1}} + (r^2+1)\BOmega^\dagger \BOmega}{2\sqrt{2}\>(r^2-1)}\> \boldsymbol{e}_1\\[7pt]
&
+e^{-it}\; \frac{2(r^2-1)  |\omega_1|^2e^{\frac{4rx}{r^2+1}} - (r^2+1)\BOmega^\dagger \BOmega}{2\sqrt{2}\>(r^2-1)}\> \boldsymbol{e}_2\\[7pt]
&
-e^{-i \frac{r^2-1}{r^2+1}\>t}\> e^{\frac{2rx}{r^2+1}} \omega_1^\ast\BOmega\>.
}
Then, we restrict ourselves to the case of $\CP^2$ and introduce the following parameterization (see~\cite{Abbott:2009um}, eq.~11)
\EQ{
\tilde\BZ= N(t,x) \Bigl( \sin\hat\xi \cos(\vartheta_2/2) e^{i\varphi_2/2} \> \boldsymbol{e}_1 + 
\sin\hat\xi \sin(\vartheta_2/2) e^{-i\varphi_2/2} \> \boldsymbol{e}_2 + \cos\hat\xi e^{i\varphi_1/2} \boldsymbol{e}_3\Bigr)\>,
}
together with $\BOmega=|\BOmega|\boldsymbol{e}_3$.
Looking at the phases, this easily leads to
\EQ{
\varphi_2= 2t\quad \text{and}\quad \varphi_1= -2\omega t\>, \quad \text{with}\quad \omega=\frac{r^2-1}{r^2+1}\>,
}
which is part of the ansatz used by AAS. In addition, they write
\EQ{
\cos\vartheta_2 =\sech \big(\sqrt{1-\omega^2}\> 2x\big)= \sech \Big(\frac{4rx}{r^2+1}\Big)\>, \qquad
\hat\xi= \frac{\pi}{2} - e(x)\>.
}
Then, our solution matches theirs provided that
\EQ{
\tan \big(\frac{\vartheta_2}{2}\big) = -\frac{2(r^2-1) |\omega_1|^2 e^{\frac{4rx}{r^2+1}} - (r^2+1)\BOmega^\dagger \BOmega}{2(r^2-1)  |\omega_1|^2e^{\frac{4rx}{r^2+1}} + (r^2+1)\BOmega^\dagger \BOmega} = -\frac{e^{\frac{4rx}{r^2+1}} -1}{e^{\frac{4rx}{r^2+1}} +1}\>.
}
This requires to fix
\EQ{
\frac{\BOmega^\dagger \BOmega}{|\omega_1|^2} =2\>\frac{r^2-1}{r^2+1}\equiv 2\omega
}
which, taking~\eqref{CentreNew} into account, amounts to fix the position of the magnon in space-time.
Notice that this choice is only possible provided that $r^2>1$.
Then,
\SP{
&\widetilde \BZ  =
-e^{it}\; \frac{e^{\frac{4rx}{r^2+1}} + 1}{\sqrt{2}}|\omega_1|^2\> \boldsymbol{e}_1 +e^{-it}\; \frac{e^{\frac{4rx}{r^2+1}} - 1}{\sqrt{2}}|\omega_1|^2\> \boldsymbol{e}_2-e^{-i \frac{r^2-1}{r^2+1}\>t}\> e^{\frac{2rx}{r^2+1}} \> \omega_1^\ast\BOmega\\[5pt]
&
\Longrightarrow |N|^2= |\omega_1|^4 \Big(e^{\frac{8rx}{r^2+1}} + 2\>\frac{r^2-1}{r^2+1}\>e^{\frac{4rx}{r^2+1}} +1\Big)\\[5pt]
&\qquad\qquad
\equiv
2 |\omega_1|^4 e^{\sqrt{1-\omega^2}\> 2x} \cosh\big(\sqrt{1-\omega^2}\> 2x\big) \big[1 + \omega\sech\big(\sqrt{1-\omega^2}\> 2x\big)\big]\>,
}
which finally leads to
\SP{
\sin^2\hat\xi=\cos^2 e(x)& = \frac{1}{|N|^2 \cos^2(\vartheta_2/2)}\>\Big(\frac{e^{\frac{4rx}{r^2+1}}+1}{\sqrt{2}}|\omega_1|^2\Big)^2\\[5pt]
&=\frac{1}{1+\omega\sech\big(\sqrt{1-\omega^2}\> 2x\big)}\>,
}
which coincides precisely with eq.~13 of~\cite{Abbott:2009um}.

\end{document}